# Growth of stylolite teeth patterns depending on normal stress and finite compaction

*Koehn, Daniel; **Renard, François; ***Toussaint, Renaud; *Passchier, Cees W.

*Tectonophysics, Institute of Geosciences, University of Mainz, Becherweg 21, 55099 Mainz, Germany, email: koehn@uni-mainz.de
**LGIT-CNRS-Observatoire, Université Joseph Fourier BP 53, F-38041 Grenoble, France & Physics of Geological Processes, University of Oslo, Norway
***Institut de Physique du Globe de Strasbourg, UMR CNRS 7516, 5 rue Descartes, F-67084 Strasbourg Cedex, France

**Abstract**

Stylolites are spectacular rough dissolution surfaces that are found in many rock types. They are formed during a slow irreversible deformation in sedimentary rocks and therefore participate to the dissipation of tectonic stresses in the Earth's upper crust. Despite many studies, their genesis is still debated, particularly the time scales of their formation and the relationship between this time and their morphology.

We developed a new discrete simulation technique to explore the dynamic growth of the stylolite roughness, starting from an initially flat dissolution surface. We demonstrate that the typical steep stylolite teeth geometry can accurately be modelled and reproduce natural patterns. The growth of the roughness takes place in two successive time regimes: i) an initial non-linear increase in roughness amplitude that follows a power-law in time up to ii) a critical time where the roughness amplitude saturates and stays constant. We also find two different spatial scaling regimes. At small spatial scales, surface energy is dominant and the growth of the roughness amplitude follows a power-law in time with an exponent of 0.5 and reaches an early saturation. Conversely, at large spatial scales, elastic energy is dominant and the growth follows a power-law in time with an exponent of 0.8. In this elastic regime, the roughness does not saturate within the given simulation time.

Our findings show that a stylolite's roughness amplitude only captures a very small part of the actual compaction that a rock experienced. Moreover the memory of the

33 compaction history may be lost once the roughness growth saturates. We also show that the
34 stylolite teeth geometry tracks the main compressive stress direction. If we rotate the external
35 main compressive stress direction, the teeth are always tracking the new direction. Finally, we
36 present a model that explains why teeth geometries form and grow non-linearly with time,
37 why they are relatively stable and why their geometry is strongly deterministic while their
38 location is random.

39 **1. Introduction**

40 Dynamic roughening of interfaces is an important research topic in many scientific
41 disciplines. A natural and spectacular example of such processes are stylolites, pairs of
42 dissolution surfaces facing each other, that are found in many rocks, mostly in limestones, and
43 which are often used for ornamental stone. A characteristic feature of stylolite interfaces is
44 their pronounced roughness with "teeth"- or "pen"-like geometries [stylus = pen, 1-5]. The
45 dark residual material that is collected within the stylolite consists mostly of clay particles [6].
46 Stylolites have mainly been described qualitatively in Earth Sciences so that dynamic models
47 of their growth and scaling properties are only now emerging [6-8]. However, the
48 development mechanism of their teeth-geometry, with peculiar square peaks (Fig. 1), is up to
49 now not understood.

50 Stylolites tend to grow perpendicular to the maximum compressive stress direction,
51 and exhibit a pronounced roughness on several scales [3]. Teeth like peaks develop with sides
52 oriented sub-parallel to the maximum stress direction (Fig. 1)[ 9]. The orientation of stylolites
53 and their teeth is commonly used by geologists as an indicator for the direction of the
54 maximum compressive stress [10-12] and the amplitude of the stylolite roughness is
55 sometimes used as a direct estimate of the compaction that the rock underwent [13]. We use
56 compaction here not only for a reduction in porosity of a rock but also for a vertical
57 shortening of the rock due to the weight of the overlying sediments. Studies on the validity of
58 these "rule of thumb" methods of compaction estimates are clearly missing, and the reason for this is probably the complexity of the development of stylolites. Since stylolites are so complex, there is also an ongoing debate in the literature about ways to classify them.

These features grow as a function of physical and chemical interactions and it is not easy to study them experimentally [14] nor in numerical models [6]. However, experiments or numerical models are needed since field observations alone cannot elucidate the dynamics of the roughening process but only represent snapshots at a given time. Early works in Earth Sciences classify stylolites according to their shape using qualitative methods [3]. However, this classification cannot successfully encompass all stylolite patterns, which can be found in numerous rock types and on a large range of scales [1-3]. Also, these classifications are often not related to any specific material parameters, scaling properties or growth conditions. More promising methods use qualitative descriptions of the stylolite roughness suggesting that they are fractal surfaces [7,15] and self-affine structures [6,8,14]. Here we present new microdynamic simulations of stylolite roughening that allow us to explore the dynamics of the stress induced roughening process in time and space, illustrate stylolite geometries that develop at different scales and propose a process-based explanation for the growth of stylolite teeth. These simulations are based on first principles of physics and chemistry, without any ad hoc phenomenological equation. It should also be underlined that contrarily to previous linearized models [6,8], the model used here allows the exploration of the fully developed structures, i.e. they take into account both, the non-linearities of the system that can be associated to large departures from flat surfaces and the solid-solid contacts occurring through the stylolite.

**2. The numerical model**

We start with an initially flat interface where dissolution can take place and the solid can only dissolve and not precipitate. This would be the case with an undersaturated fluid facing the solid. This flat interface may represent an initial "anticrack", that is assumed to represent the stylolite at an early stage of formation [16,17]. The model is two-dimensional

85 and it assumes that everything that dissolves along the interface is transported instantaneously
86 in the fluid out of the system, i.e. that diffusion in the fluid pocket is not limiting the process,
87 but happens at a much faster time scale than the dissolution itself. Accordingly, the fluid has a
88 constant solute concentration. The model is part of the modelling environment "Elle" [18].
89 The setup of the model is as follows: two solids are pressed together with a confined fluid
90 layer in between them (Fig. 2). The right and left hand side of the model are fixed through
91 elastic walls whereas the lower and upper walls of the model are moved inwards at a constant
92 displacement rate. The solid is made up of small particles that are connected via linear elastic
93 springs along a hexagonal lattice. These elements can represent either a single grain or a pack
94 of smaller grains. At a scale larger than the grains, this network behaves in a classical
95 elastostatic way.

96 *2.1 Thermodynamics and kinetics of stylolite dissolution*

97 Dissolution of the solid takes place in small steps dimensioned in such a way that one
98 element of the solid is dissolved at every step. Dissolution follows a simple linear rate law

99 $$D = kV\left[1 - \exp\left(-\frac{\{\Delta\psi + \Delta\sigma_n\}V}{RT}\right)\right], \qquad (1)$$

100 where $D$ is the dissolution velocity of the interface (m s$^{-1}$), $k$ a dissolution kinetics rate
101 constant (mol m$^{-2}$ s$^{-1}$), $V$ the molecular volume of the solid (m$^3$ mol$^{-1}$), $R$ the universal gas
102 constant (J mol$^{-1}$ °K$^{-1}$), $T$ the temperature (°K), $\Delta\psi$ (J m$^{-3}$) the changes in Helmholtz free
103 energy of the solid during dissolution of a solid element, and $\Delta\sigma_n$ (Pa) the normal stress
104 gradients along the interface [further details of the derivation are given in 19,20]. The
105 Helmholtz free energy takes into account the variations in elastic and in surface energy.

106 Changes in surface energy are calculated from the curvature of the interface. The
107 surface energy ($E^s$) per area unit around a single element located at the interface can be
108 expressed as

109 $$E^s = \frac{\gamma}{\rho}, \qquad (2)$$

110 where $\gamma$ is the surface free energy and $\rho$ is the local radius of curvature of the interface. The

111 local curvature is determined using the two neighbours of each element along the interface.

112 The sign of the curvature $\rho$ is such that it is driving dissolution in the rate law when the solid

113 is convex towards the fluid, and precipitation when it is concave. We perform an average

114 across the interface using the local surface energies of elements and those of their neighbours

115 (up to n = 40 in each direction) with the expression

116 $$E_i^{avs} = \frac{1}{C} \sum_{h=1}^{n} \left( \frac{1}{(2h+1)^2} \sum_{j=0}^{2h+1} E_{i+j-h}^s \right), \qquad (3)$$

117 where $E_i^{avs}$ is now the average surface energy for element $i$, $E_i^s$ the local surface energy for

118 element i, and $E_{i+j-h}^s$ surface energies of neighbouring elements along the interface, and

119 $C = \frac{\pi^2}{4} - \frac{17}{9} \approx 0.578512$ is a normalization factor that ensures that the average surface

120 energy on all sites is preserved in this averaging procedure. The local surface energies are

121 divided by the sum of added elements and their maximum distance (divided by the initial

122 radius of curvature of a particle). This averaging procedure amounts to consider a coarse-

123 grained surface energy, at the scale of a few elements, and allows to avoid artefacts that could

124 arise from the discreteness of the model. In this way, a singular local term $E_i^s$ obtained at a

125 corner of the interface is smoothened for $E_i^{avs}$ over a neighbourhood of a few grains. This

126 procedure is commonly used in computational physics to avoid discreteness artefacts [21,22].

127 The above expression is equivalently represented as a discrete convolution operation

128 $$E_i^{avs} = \sum_{-\infty < j < \infty} f_j E_{i+j}^s. \qquad (4)$$

129 with a tent function shown on Fig. (2a),

130 $$f_j = \frac{1}{C}\sum_{k=|j|}^{\infty}\frac{1}{(2k+1)^2} \text{ for } j \neq 0, \text{ and } f_0 = f_1. \qquad (5)$$

131 Keeping the terms up to n = 40 in this sum, i.e. all terms in the sum of Eq. (4) for $|j| < 40$, we

132 are left with an error of only 2% in the tail of the weight function. The value of the

133 normalization constant C is obtained by requiring that the average coarse grained surface

134 energy is equal to the average local surface energy, i.e. it is set up by the condition that

135 $$\sum_{-\infty < j < \infty} f_j = 1. \qquad (6)$$

136

137 *2.2 Mechanics*

138 The elastic energy and the normal stress at the interface are determined using a lattice

139 spring model for the solid where elements are connected by linear elastic springs. We assume

140 that the solid deforms only elastically without internal plastic deformation, except for the

141 irreversible dissolution events happening at the modelled interface. Stresses in the solid are

142 determined using an over-relaxation algorithm where elements of the model are moved until a

143 new equilibrium configuration is found. The equilibrium configuration is defined by a given

144 relaxation threshold. The net force ($F_i$) acting on an element $i$ at position $x_i$ is

145 $$F_i = \sum_{(j)} \kappa \left(|x_i - x_j| - l\right) v_{i,j} + f_p, \qquad (7)$$

146 where the sum is over all neighbours $j$, $\kappa$ is a spring constant, $l$ is the equilibrium distance

147 between elements $i$ and $j$, $v_{i,j}$ is a unit vector pointing from $j$ to $i$ and $f_p$ is an external force

148 like the repulsion from a wall.

149 The elastic energy ($E^{el}$) is directly evaluated at a node $i$ as

150 $$E^{el} = \frac{1}{4V}\sum_{(j)} \kappa \left(|x_i - x_j| - l\right)^2, \qquad (8)$$

151 where $V = \sqrt{3}/2 l^2$ is the volume of an elementary cell. It can also be determined from the

152 strain tensor ($u_{ik}$) that is calculated from the lattice spring model with the expression [23]

153 $$E^{el} = \frac{1}{2} \lambda_1 \left( \sum_i u_{ii} \right)^2 + \lambda_2 \sum_{i,k} (u_{ik})^2, \qquad (9)$$

155 where $\lambda_1$ and $\lambda_2$ are the Lamé constants. We use the Einstein convention with summation
156 over repeated indices. Differences in elastic energy in equation (9) refer to differences
157 between a stressed and a non-stressed element. The Lamé constants are set up by the spring
158 constant, the lattice constant and the hexagonal lattice configuration, i.e. $\lambda_1 = \sqrt{3}\kappa/(2l)$ and
159 $\lambda_2 = \sqrt{3}\kappa/(4l)$, or equivalently, the Young modulus is $K = \lambda_1 + 2\lambda_2/3 = 2\kappa/(\sqrt{3}l)$, and the
160 Poisson ratio is $\nu = \lambda_1/(2\lambda_1 + 2\lambda_2) = 1/3$ [24].
161 Finally the normal stress at the interface is determined from the repulsion of the two solids at
162 the interface where the repulsion only contains normal components, assuming that a fluid film
163 with no shear stress exists at the interface [25]. In order to calculate changes in normal stress
164 along the interface we calculate an average of the normal stress across the whole interface and
165 define differences in stress as the local normal stress minus the average normal stress.
166     The simulation flow is as follows:
167 - First the outer walls are moved in a given time step to stress the system.
168 - Once the upper and lower solids meet at the interface stress builds up. The rate law
169     (eq. 1) is then used to calculate if elements can dissolve in the given time step. If not
170     the system is stressed again until the first element can dissolve.
171 - Once elements dissolve they are removed one at a time and the stress is calculated
172     again. The time that is used up by the dissolution of a single element is averaged to be
173     the time it takes to dissolve that element completely divided by the system size
174     (number of elements in the x-direction). Dissolution of elements proceeds until the
175     given time is used up and another deformation step is applied. Using a desktop
176     workstation, each simulation lasts between 10 and 30 days; the stress relaxation being
177     the most time consuming part.

*2.3 Simulation parameters and disorder*

The parameters used in the simulation should mimic those of the natural example (fig. 1). For our idealized model we use a rock mainly made up of calcite with a molecular volume of 0.00004 m$^3$/mol, a Young's Modulus of 80 GPa, a Poisson's ration of 0.33 (given by the model configuration), a surface free energy of 0.27 J/m$^2$, a temperature of 300 K and a dissolution rate constant of 0.0001 mol/(m$^2$s) [8]. In addition, the displacement rate of the upper and lower boundaries is fixed at a constant value corresponding to strain rates of compaction between $10^{-10}$ and $10^{-12}$ s$^{-1}$ (see fig. 2).

In order to introduce heterogeneities to the system a bimodal variation is set on the dissolution rate constants of the elements. The heterogeneity in the system is set such that 5% of all elements have a dissolution rate constant that is half the rate constant of the other elements, i.e. they dissolve slower and pin the surface. The initial spatial distribution of the rate constants is set using a pseudorandom routine resulting in a probability of 5% of particles dissolving more slowly, picked independently for each site. Using this procedure, a spatial heterogeneity, also called quenched noise, is introduced in the initial rock.

## 3. Results

*3.1 Simulated stylolites and comparison with natural data*

At first we compare a simulated stylolite directly with a natural example. Figure 3 shows a simulation of a roughening stylolite in a model that is 400 elements wide and that has an absolute horizontal size of 40 cm. One element in the model then has a diameter of 1 mm and may represent a single grain in a natural rock. We can compare the simulation with the natural stylolite shown in figure 1, especially with the inset. The simulated stylolite has a width that is about 2/3 of the width of the natural stylolite in the inset (see the hammer for the scale).

203    The simulated stylolite develops in 8000 years at a compaction rate of $3 \times 10^{-12}$ s$^{-1}$ and

204    has compacted by 25.6 cm or 64% (original height of the simulated box was 40 cm). Both

205    simulated and natural stylolites are visually very similar. They both produce pronounced teeth

206    with smaller scale roughness in between and on top of the teeth. The height of the teeth (up to

207    about 8 cm) and their width are comparable in the natural and the simulated stylolite implying

208    that both have similar scaling properties. The grain size of the simulated example (1mm) is

209    larger compared to natural rocks, mainly because we are limited by the calculation time of the

210    model.

211    *3.2 Initiation of the stylolite roughness by interface pinning along heterogeneities*

212    The stylolite roughness is initiated by elements that dissolve slower. If the model

213    contains no heterogeneity the interface will only become rough on the scale of single elements

214    and remain flat on the larger scale. This is because surface energies and elastic energies are

215    minimized when the surface is flat. Therefore, both of these energies will prevent the surface

216    to roughen and the stylolite to grow [8]. Once the system contains heterogeneity, slower

217    dissolving elements continuously pin the surface and thus make it rougher (Fig. 4).

218    Dissolution takes a longer time to destruct a roughness that is pinned by slower dissolving

219    elements than to flatten an interface with no variation. The roughness is not stable but very

220    dynamic in time since an increasing amount of more slowly dissolving elements are pinning

221    the interface while the solid progressively dissolves. However, the slower dissolving elements

222    themselves may dissolve as well if the roughness produces locally very high surface energies

223    due to a strong curvature of the interface or high elastic energies due to stress concentrations

224    or if two slower dissolving elements meet on opposing interfaces. Dissolution of pinning

225    elements will then reduce the roughness again.

226    An example of the development of the roughness in a simulation and the effects of

227    interface pinning is illustrated in figure 4. Figure 4a shows the initial random distribution of

228    slower dissolving elements (small dark spots). This heterogeneity is frozen into the system at

229 the beginning and does not change during a simulation (quenched noise). The interface where
230 the stylolite is initiated is shown as a black line. While the solid dissolves, pinning elements
231 are progressively being collected within the interface (small white spots, fig. 4b to d).
232     Two distinctively different patterns develop during the pinning of the interface. On
233 one hand, one or a couple of elements pin very locally and produce local spikes. On the other
234 hand, larger parts of the interface may be pinned between two elements that are further apart.
235 These second pattern generates the teeth geometries, typical of stylolites, with teeth having
236 various widths.
237     Figure 4e illustrates three different cases of pinning schematically. Single pinning
238 elements produce spikes whereas two pinning elements that pin from the same side produce
239 teeth. The structures grow fastest when elements pin from opposing sides. The interfaces in
240 figures 4c and 4d illustrate the teeth-forming processes presented in figure 4e: the interface is
241 made up of single pinning elements, larger teeth where groups of elements pin and steep
242 interfaces where elements pin from opposing sides. The surface structure changes when new
243 pinning elements are collected within the stylolite and when pinning elements are destructed.
244 The amplitude of the stylolite grows during these processes (from figure 4b to d) and the
245 wavelength of the interface is also evolving. Small wavelengths can grow very fast (figure 4c)
246 whereas the larger wavelengths need longer time to develop (figure 4d).
247     Figure 5 shows the evolution of the growing roughness of two stylolites through time
248 in 3D diagrams where the x-axis shows the amount of elements in the x direction, the y-axis
249 corresponds to the time in model time-steps (here one step corresponds to 20 years) and the z-
250 axis shows the non-dimensional height of the stylolite. The parameters for the two stylolites
251 shown in figure 5a and 5b are the same except for the absolute length (and height
252 respectively). The stylolite shown in figure 5a has an x-dimension of 0.4 cm whereas the
253 stylolite shown in figure 5b has an x-dimension of 40 cm.

254	The differences in absolute initial system size $L$ have an effect on the dominance of
255	elastic versus surface energies during the roughening process. Surface energies become
256	increasingly more important towards smaller scales. This means that the stylolite in figure 5a,
257	which is relatively small ($L$ = 0.4 cm) with a small grain size of 10 µm, is dominated by
258	surface energies so that elastic energies only play a minor role. The stylolite shown in figure
259	5b however is relatively large ($L$ = 40 cm) with a large grain size of 1 mm so that the elastic
260	energy dominates the roughening process and surface energies only play a minor role.
261	When comparing figures 5a and b (note that the z-axis scales differently, one unit
262	corresponds to 0.4 cm in a) and to 40 cm in b)), one observes that the roughness forms better
263	developed teeth with steep sides in the case of the larger stylolite (fig. 5b). The roughness of
264	the smaller stylolite (fig. 5a) is not growing smoothly but is disrupted quite often and
265	produces neither large amplitude nor well-developed teeth. In addition, the larger stylolite
266	(Fig. 5b) grows progressively while the small stylolite (Fig. 5a) shows an initial increase in
267	roughness that is followed by strong fluctuations in time, where the average roughness
268	amplitude remains more or less constant. These roughness evolutions imply that pinning
269	elements are destroyed when surface energy is high because of very high curvatures of spikes.
270	The interface of the surface energy dominated stylolite therefore contains no larger spikes or
271	teeth and is quite dynamic. Elastic energy on the other hand does not destroy spikes easily.
272	Therefore well-developed teeth structures tend to arise in larger stylolites at the outcrop scale
273	when surface energy is relatively unimportant, whereas we would expect to find less well-
274	developed teeth but rounder structures on the scale of a thin-section, where surface energy is
275	important.
276	In order to explore the evolution of the roughness amplitude with time, we use signal-
277	processing tools from statistical physics [26], as illustrated in the next section.
278	*3.3 Growth of the roughness with time*

279     The dynamics of a surface roughening process can be described by some basic scaling

280 laws that are the same or at least very similar for different interfaces and surfaces (as e.g., gas-

281 fluid interface motion in non-saturated porous media, propagation of flame fronts, atomic

282 deposition processes, bacterial growth, erosion or dissolution fronts, contact line motion

283 biphasic fronts over disordered pinning substrates, interfacial crack fronts). These laws,

284 discovered by statistical physicists, describe how the amplitude of the roughness grows non-

285 linearly with time, following power laws [26-29].

286     First we have to define an average value for the amplitude of the roughness of our

287 numerical stylolites for each time step. We use the root mean square method to determine the

288 average width of the stylolite roughness following [26]

289
$$w(L,t) \equiv \sqrt{\frac{1}{L}\sum_{i=1}^{L}\left[h(i,t)-\overline{h}(t)\right]^2}, \qquad (10)$$

290 where $w$ is the interface width as a function of system size $L$ and time $t$, $h$ is the height of

291 point $i$ on the interface at time $t$ and $\overline{h}$ the average height of the interface at time $t$. This

292 function gives an average width of the interface for each time step $t$ and therefore

293 characterizes the growth of the roughness. In our simulations, the system size $L$ is defined as

294 the number of elements in the x-direction. We use model sizes of 200 and 400 elements in the

295 x-direction. The average height of the interface is defined as

296
$$\overline{h}(t) \equiv \frac{1}{L}\sum_{i=1}^{L}h(i,t). \qquad (11)$$

297     Statistical physics scaling laws [26] have demonstrated that, in many stochastic

298 interfacial systems, roughening interfaces grow following a power law in time with a so-

299 called growth exponent $\beta$. This is described by

300
$$w(L,t) \sim t^{\beta}, \qquad (12)$$

301 where the interface width $w$ is proportional to time $t$ to the power $\beta$, for a given system size $L$.

302 If $\beta = 1.0$, the interface grows linearly with time, if $\beta$ is smaller than 1.0 the interface growth slows down with time. Normal diffusion processes are characterized by $\beta=1/2$, anomalous diffusion processes by $\beta \neq 1/2$, and $0 \leq \beta \leq 1$. In addition, for most stochastic interfacial systems the width of roughening interfaces saturates after a critical time $t_{crit}$. This time increases with the system size $L$.

When modelling the growth of stylolites, one can expect two scaling regimes in time, first a power law up to time $t_{crit}$ followed by a regime where $w$ remains constant (Fig. 6a). In order to study the dynamics of the roughening process one constructs diagrams of $\log_{10}(w)$ against $\log_{10}(t)$. The increase in width of the interface roughness should follow a straight line where the slope of the line gives the growth exponent $\beta$. After a critical time $t_{crit}$ the roughness saturates and the slope vanishes to zero (Fig. 6a).

We studied three simulations (Fig. 6b-d) with this method, where the simulation shown in figure 6b has a system size $L = 0.4$ cm, the simulation shown in figure 6c has $L = 4$ cm and the one shown in figure 6d has $L = 40$ cm. Figure 6b and 5a and figure 6d and 5b show the same simulations, respectively. The simulated stylolite shown in figure 6b shows the expected behaviour with two scaling regimes, the roughness first grows with a roughness exponent of 0.5 and saturates after 2500 years, where it remains constant close to a value of 50 µm. Going back to figure 5a where the growth of the same stylolite is illustrated in 3d, the roughness saturates after 2500 years, which corresponds to model time step 125. Figures 6c and d show only the first scaling regime where the increase in roughness amplitude $w$ follows a power law, but $w$ never saturate. That means probably that these two simulations (Fig. 6c, d) did not reach the critical time needed for the roughness to saturate. Taking a look at figure 5b where the growing roughness of the stylolite shown in figure 6d is illustrated, the roughness width still grows i.e. is not yet saturated.

The three stylolites shown in figure 6b-d seem to have different critical times when the roughness saturates but also their growth exponents vary. $\beta$ increases from the smallest simulation (fig. 6b) with a value of 0.5 through the medium-sized simulation with a value of

329  0.54 to the largest simulation with a value of 0.8. These differences may reflect differences of
330  growth regimes that are dominated by either surface energy or by elastic energy, in analogy to
331  the discussion on figures 5a and b. The surface energy dominated growth regime (small
332  stylolite, fig. 6b) has a growth exponent of 0.5 and saturates relatively early. The stylolite
333  shown in figure 6c is intermediate, the growth exponent is still small with 0.54 but the
334  roughness does not saturate within the simulated time. The larges stylolite (fig. 6d) has a
335  significantly larger growth exponent of 0.8, does not saturate, and represents the elastic
336  energy dominated regime.

337  In terms of natural stylolites the above-mentioned values demonstrate that small
338  stylolites that grow within the surface energy dominated regime grow as slow as a diffusive
339  process (exponent of 0.5) and saturate early so that compaction estimates are almost
340  impossible. However, stylolites that grow within the elastic energy dominated regime grow
341  faster (exponent 0.8) and do not seem to saturate after 8000 years. They can thus capture part
342  of the compaction of the rock even though their growth is non-linear, and slows down with
343  time. Consequently, rather than using a rule of thumb as a direct proportionality between $A$,
344  the amplitude of compaction displacement accommodated for around a stylolite, and the
345  stylolite amplitude $w$, the nonlinear power-law observed for this process where the imposed
346  displacement is linear in time can be utilized to be stated as $(w/l) \sim (t/t_0)^\beta \sim (A/l)^\beta$, with
347  $l$ the relevant physical length, which is here the grain size.

348  Conversely, for large enough stylolites, as long as the critical saturation time has not
349  been reached at the observed scale, the relationship between the total compaction
350  displacement A and the stylolite amplitude should be of the type

351 $$A \sim (w/l)^{1/\beta} l. \qquad (13)$$

352  This law should hold until the critical saturation time is reached. Using the proposed scaling
353  law for the large simulation ($L = 40$cm) we obtain the right relation with a slope of 6.6 (Fig.
354  7) using the root mean square width and a grain size of 1mm. The stylolite amplitude may finally saturate, but this seems not to happen in 8000 years for the case of the 40 cm long stylolites that we have studied.

*3.4 The teeth structures and their relation with the main compressive stress direction*

After characterizing the dynamics of the growth process, we now focus on the orientation of the stylolite teeth. We have demonstrated so far that the teeth mainly develop in the regime where elastic energy is dominant so that well-developed teeth form in the stylolite simulation with a system size $L = 40$ cm. The steep sides of the teeth are thought to develop parallel to the main compressive stress and the top of the teeth is thought to be oriented perpendicularly to that stress direction.

We test this hypothesis that the teeth track the orientation of the largest principal stress. For the test, we choose again a model with an initially flat dissolution surface that is oriented horizontally with respect to the simulation box but we rotate the compression direction so that the stress field rotates. Theoretically, the teeth should follow the rotated stress field so that one can evaluate the direction of the main compressive stress using the teeth orientation. We use simulations with a system size of 200 elements in the x-direction with an absolute size of $L = 10$ cm so that we are in the elastic regime.

The results of the 5 different simulations are shown in figure 8 a-e where the orientation of the compression direction is shown on the left hand side. Indeed, the teeth do follow the stress directions in all the examples. The sides of the teeth are the best stress direction indicators but even the tops of some teeth tend to orient themselves with respect to the stress. The most extreme example is shown in figure 8e where the compression direction has a very low angle with respect to the initial horizontal heterogeneity. The initially horizontal surface has vanished and a stair-step geometry develops with steps that are oriented perpendicular with respect to each other. One set of surfaces corresponds to the sides of the teeth and the other set to the top surface of the teeth.

380    These simulations demonstrate that stylolite teeth indeed track the direction of the
381    main compressive stress and can therefore be used as stress indicators in natural rocks.
382    *3.5 Model for the deterministic orientation of stylolite teeth*
383    Our simulations have shown that well-developed stylolite teeth only tend to develop in
384    the elastic regime (Fig. 5). In order to understand the relation between the compaction
385    direction and the orientation of the stylolite teeth, we consider the stress distribution at the
386    interface and the finite compaction. The stress field at the interface for a given time step is
387    directly controlling dissolution. However, just observing the stress field across a stylolite for a
388    single time step is not enough since dissolution and thus a change in the geometry of the
389    interface influences the stress back, dissolution and stress being highly coupled. Therefore we
390    also have to consider the whole stress history and the finite compaction across the interface.
391    Going back to figure 4, one can observe that the location of the teeth is random
392    because it depends on the quenched noise in the background. The second simple observation
393    is that the finite compaction is directly recorded by the pinning of particles (Fig. 4). As long
394    as they pin the interface, they move in the direction of the compaction, which is parallel to the
395    far field compressive stress for a simple homogeneous rock. This scenario does not change
396    when the compaction direction and thus the main compressive stress is not perpendicular to
397    the initial interface (Fig. 9a). Pinning particles still record the rotated compaction direction
398    because they move in that direction. If two pinning elements are close but on opposing sides
399    of the interface they move in opposite directions and a perfect side of a tooth develops
400    recording the compaction direction. If this direction is parallel to the far field main
401    compressive stress, the stylolite also records this stress direction.
402    Looking at the movement of a whole interface (Fig. 9b), one can see that the total amount that
403    the interface may move perpendicularly to itself depends on its orientation with respect to the
404    compaction direction. An interface that is oriented perpendicularly to the compaction
405    direction can have the largest fluctuations and thus produce a roughness with the largest amplitude. An interface that is oriented at a smaller angle to the compaction direction will not develop importantly in the direction perpendicular to itself. If particles are pinning this interface, it may actually develop steps. The most extreme case is an interface that is oriented parallel to the compaction direction. This interface shows no fluctuations and thus cannot develop a roughness. It cannot even develop steps because particles cannot pin this interface. The interface is very stable and can only act as a transform fault. Therefore the sides of teeth, which reflect such an interface, are very stable and the teeth geometry itself is a natural consequence of the compaction direction (Fig. 9c). Strongly pinning particles are not even necessary for the development of these geometries. Some pinning or a variation in dissolution across the interface is necessary for a roughening of the interface, but once this roughness is initiated and for large scales where the surface energy is not dominant, the typical teeth geometries develop and stay stable without much flickering, i.e. they are strongly deterministic.

Looking at the stress history, we also show that the main compressive stress close to the stylolite is parallel to the main compaction direction. An interface perpendicular to the main compaction direction has a significantly normal stress component and experiences dissolution. For an interface parallel to the main compressive stress, dissolution will only relax the second principle stress that builds up in a laterally confined system. The second principle stress will vanish to zero and in the extreme case a hole can develop. Compaction itself cannot fill the hole because the compaction has the wrong direction. The only possibility to build-up stress again is a flow of material into the hole or a collapse of the teeth. Looking at the stress history, the main compressive stress close to the stylolite and the direction of the finite compaction are parallel. Indeed, the stylolite teeth track this direction.

**4. Discussion**

In this discussion we first focus on several assumptions used in the simulations. The focus of these simulations is the roughening process of an initially smooth interface, not an investigation of the reason why the dissolution is localized. Therefore we use an initially flat interface were dissolution starts and restrict dissolution to this stylolite surface. Some observations in natural rocks support such a scenario: First, some stylolites do initiate from mica-rich layers. Next, some stylolites with oblique teeth do exist. Stylolites with oblique teeth are most probably initiating from an interface that was oblique with respect to the main compressive stress direction when the stylolite roughness started to grow. However, surely not all stylolites start from given heterogeneities but may localize due to geochemical self-organization [30] or anti-cracking [16]. This is beyond the focus of this current manuscript.

In our simulations, we use a simple description of the noise that initiates the roughening process. As mentioned earlier, the noise is only chemical (a variation in dissolution constants), is set on single particles that have the same size and is distributed with a bimodal distribution. In a real rock, elastic properties and surface energy may also vary, the noise may be on the grain scale or at the scale of smaller heterogeneities and the distribution may be more complicated than bimodal. It is not clear how variations in these parameters may affect the stylolite growth and what kind of noise is present in a real rock. The nature of the noise may influence pinning of the surface and thus may influence the growth and structure of the stylolite [31]. However, since our simulations can reproduce natural stylolites quite realistically, one can argue that the exact nature of the noise is less important than the effects of elastic energy, surface energy and stress.

Another interesting question is how a stylolite grows when the finite strain is rotating. We have demonstrated that stylolite teeth develop by a particle pinning process at the interface and track the direction of the finite compaction. However, this is not necessarily true if the finite strain is rotating because particles at the interface only pin the interface from the moment when they meet the interface to the moment when they dissolve. This means that they are only recording parts of the finite compaction direction. If the stylolite grows in a simple shear dominated rock, the direction of the pinning particles should initially record the

458  direction of the compressive ISA (incremental stretching axes) and then the teeth should
459  rotate. Therefore young teeth do really track the compressive ISA and thus probably also the
460  direction of the compressive stress whereas old teeth rotate. However, if the roughness is
461  dynamic and the stylolite constantly changes its shape, it can always grow new teeth and old
462  teeth will disappear so that the direction of the compressive ISA and the compressive stress
463  are recorded.

464  **5. Conclusions**

465  We have developed a numerical model that can successfully reproduce the roughening
466  of stylolites. The numerical stylolites are very similar, if not identical, to natural stylolites. We
467  propose that the growth of the stylolite roughness is induced by pinning particles, that
468  produce a complex interface that evolves dynamically through time. Two different regimes
469  can be separated, a small-scale regime where the roughness fluctuates significantly and a
470  large-scale regime where well-developed teeth patterns grow. The small-scale regime is
471  dominated by surface energy whereas the large-scale regime is dominated by elastic energy.
472  Scaling laws characterizing the dynamic growth of the stylolite roughness as a function of
473  time are proposed. These laws show that the roughness grows in two successive regimes in
474  time, a first regime where the growth follows a power law and a second regime where the
475  roughness growth saturates. These findings are essential for compaction estimates using
476  stylolites, the roughness growth is non-linear in time, slows down with time and may even
477  saturate. A saturated stylolite looses its memory for compaction completely and cannot be
478  used for total compaction or strain rate estimates. We also show with our simulations that the
479  teeth of stylolites do really follow the main compaction direction and may thus indicate the
480  direction of the maximum compressive stress in a homogeneous rock. We show that this
481  strong deterministic geometry of the teeth is a consequence of pinning particles that move in
482  the direction of the finite compaction and of the local stress history at the stylolite interface.
483  Summarizing, the geometry of stylolite teeth can be used by geologists to estimate the direction of the finite compaction or the main compressive stress, but absolute compaction estimates are difficult to perform and may strongly underestimate the real values. In the regime preceding saturation, we utilize the observed nonlinear growth of the roughness amplitude to propose a refined estimate of absolute compaction, Eq. (13), based both on the stylolite roughness amplitude and the size of the dissolving grains (Fig. 7). This can in general be used as a lower bound of total compaction.

**Acknowledgements:**

DK acknowledges funding by the MWFZ Center of Mainz University and the DFG (KO 2114/5-1). This is contribution No. 37 from the Geocycles Cluster funded by the state of Rheinland-Pfalz.

**Figure captions**

Figure 1. Natural stylolites on a limestone quarry surface from the Burgundy area, France. On the large scale the stylolites are planar structures whereas on the small scale they show a pronounced roughness (see inset). Note the steep "teeth" like patterns in the inset.

Figure 2. a) weight function used in the coarse graining procedure to evaluate the surface energy. b) setup of the numerical model. Side-walls are fixed whereas the upper and lower walls are moved inwards to compact the system. The rock contains two kinds of particles that dissolve at two different rates. This heterogeneity in the dissolution rate represents an initial quenched noise in the rock.

Figure 3. Simulated stylolite with a width of $L = 40$ cm (1:1). Note similarities between the natural stylolite (in Figure 1) and the simulated stylolite. Both develop a roughness on different scales and well-developed square teeth structures.

Figure 4. Particles pinning along the interface. (a) Initial setup of a simulation with the initial interface in the centre and the quenched noise in the background (particles that dissolve twice slower are dark). (b)-(d) Progressive growth of the roughness and pinning of the interface (particles that pin are white). (e) Structures that develop during pinning: spikes, teeth that are pinned in one direction and teeth that are pinned in two directions.

Figure 5. 3D diagrams showing the growth of two stylolites. (a) Small stylolite that grows in the surface energy dominated regime (x dimension is $L = 0.4$cm). The growth of the roughness slows down relatively fast and saturates. The growth is very dynamic so that pinning particles are dissolved relatively fast. (b) Large stylolite that grows in the elastic

energy dominated regime (x dimension is $L = 40$ cm). The stylolite roughness grows continuously and develops well-developed teeth.

Figure 6. Log/log diagrams showing the growth of the roughness amplitude $w$ (one unit = 0.004 m) against time (t, model steps). (a) Schematic diagram illustrating the two regimes that should develop [26]. Regime I shows a growth that follows a power law in time with a growth exponent $\beta$ until a critical time. Regime II is characterized by a saturation of the growth so that the roughness amplitude remains constant. (b) Small stylolite ($L = 0.4$cm) that shows both regimes, a power law growth with an exponent of 0.5 in regime I (as slow as a diffusive process) and a crossover regime II where the roughness saturates. (c) Medium sized stylolite ($L = 4$ cm) that shows only the regime I with a power law growth with an exponent of 0.54. (d) Large stylolite ($L = 40$ cm) that shows only regime I with a power law growth with an exponent of 0.8.

Figure 7. Proposed scaling relation between the amplitude of compaction around a stylolite ($A$) and the mean width of the stylolite roughness ($w$). $l$ corresponds to the grain size and $\beta$ is the predetermined growth exponent (0.8 in this case). Plot shows the scaling relation for the large stylolite simulation (40cm long stylolite) using equation 13. Theoretically this scaling relation with the determined prefactor of 6.6 can be used to estimate compaction from real stylolites.

Figure 8. Stylolite teeth directions track variations of the compression direction. The orientation of the initial interface is given at the top and the direction of the compaction is illustrated at the left hand side. The simulations show that the orientation of the teeth is strongly deterministic and follows the compaction direction.

Figure 9. Model for the development of stylolite teeth. (a) Development of oblique teeth due to pinning particles that record the relative movement of the rock on each side of the interface. (b) Possible fluctuations that may develop on interfaces with different orientations with respect to the main compaction direction. Strongest fluctuations appear on interfaces that are oriented perpendicular with respect to the main compaction direction (arrows). Interfaces that are parallel to the main compaction direction do not show fluctuations. (c) Illustration of these fluctuations on a natural stylolite. The sides of teeth are relatively stable, show no fluctuations and may act as transform faults. Plateaus of teeth are interfaces that show largest fluctuations.

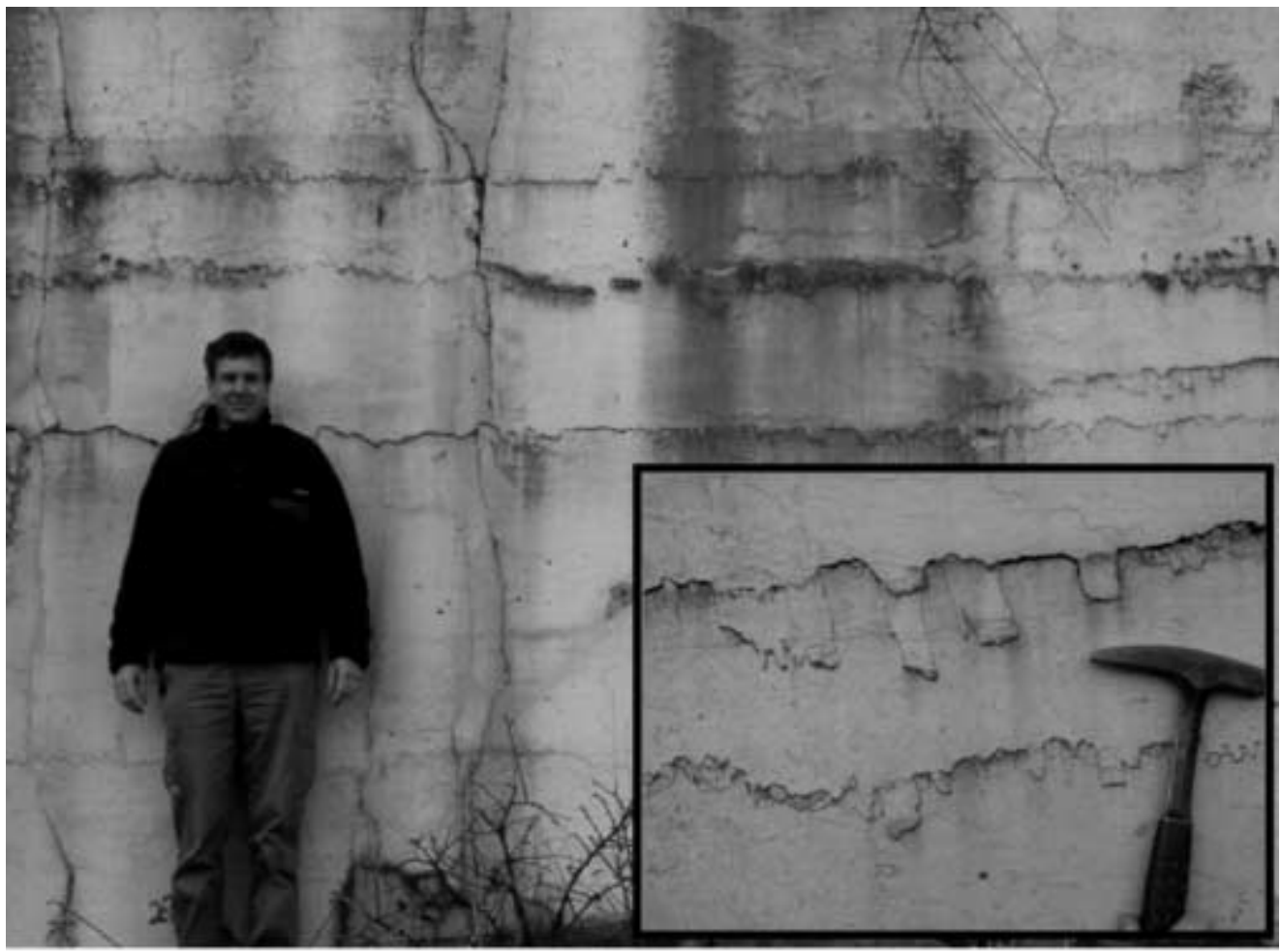

Fig. 1

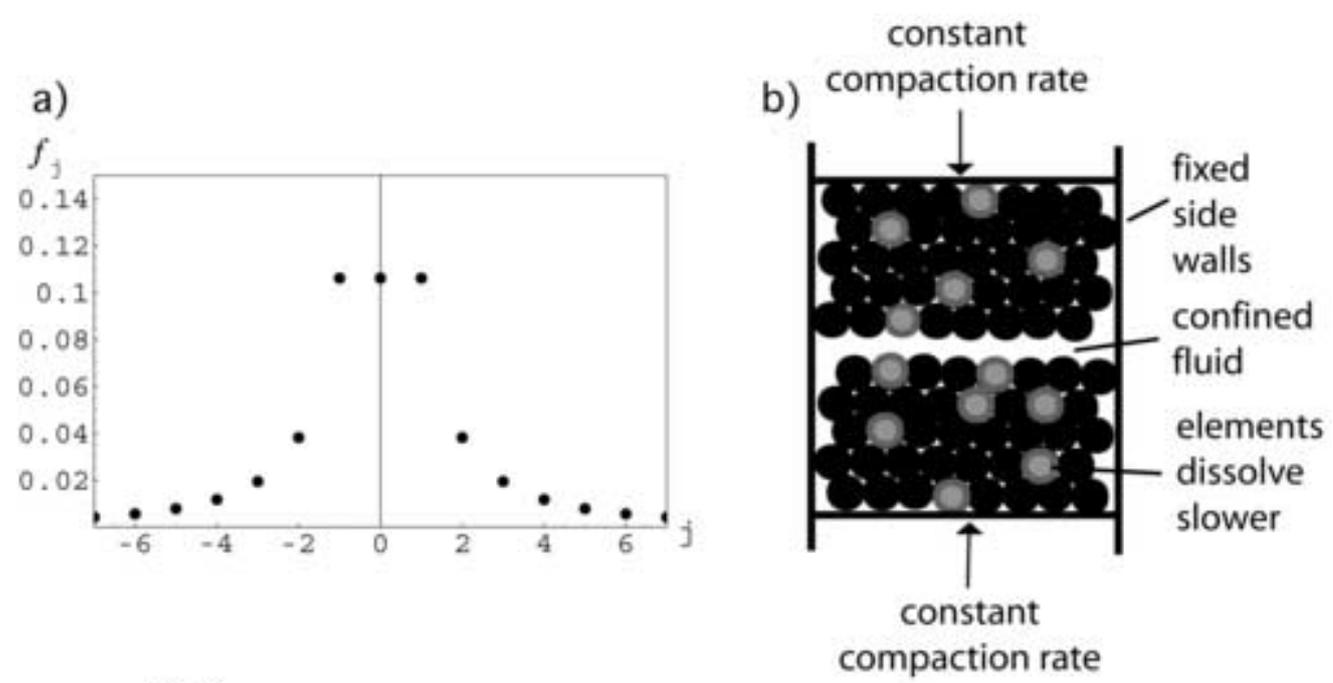

Fig. 2

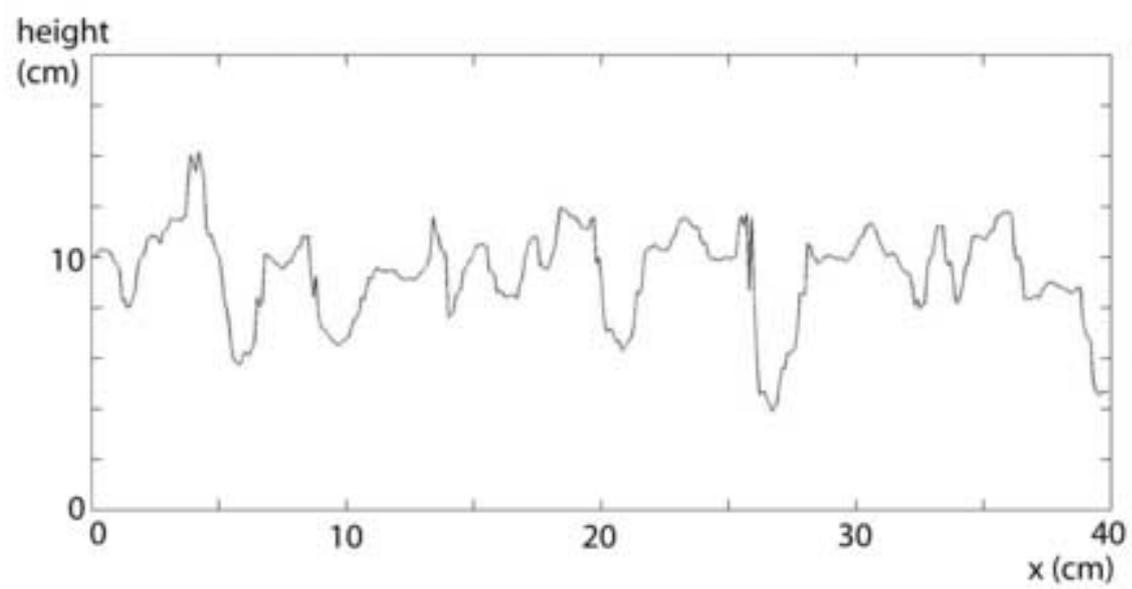

Fig. 3

Figure
Click here to download high resolution image

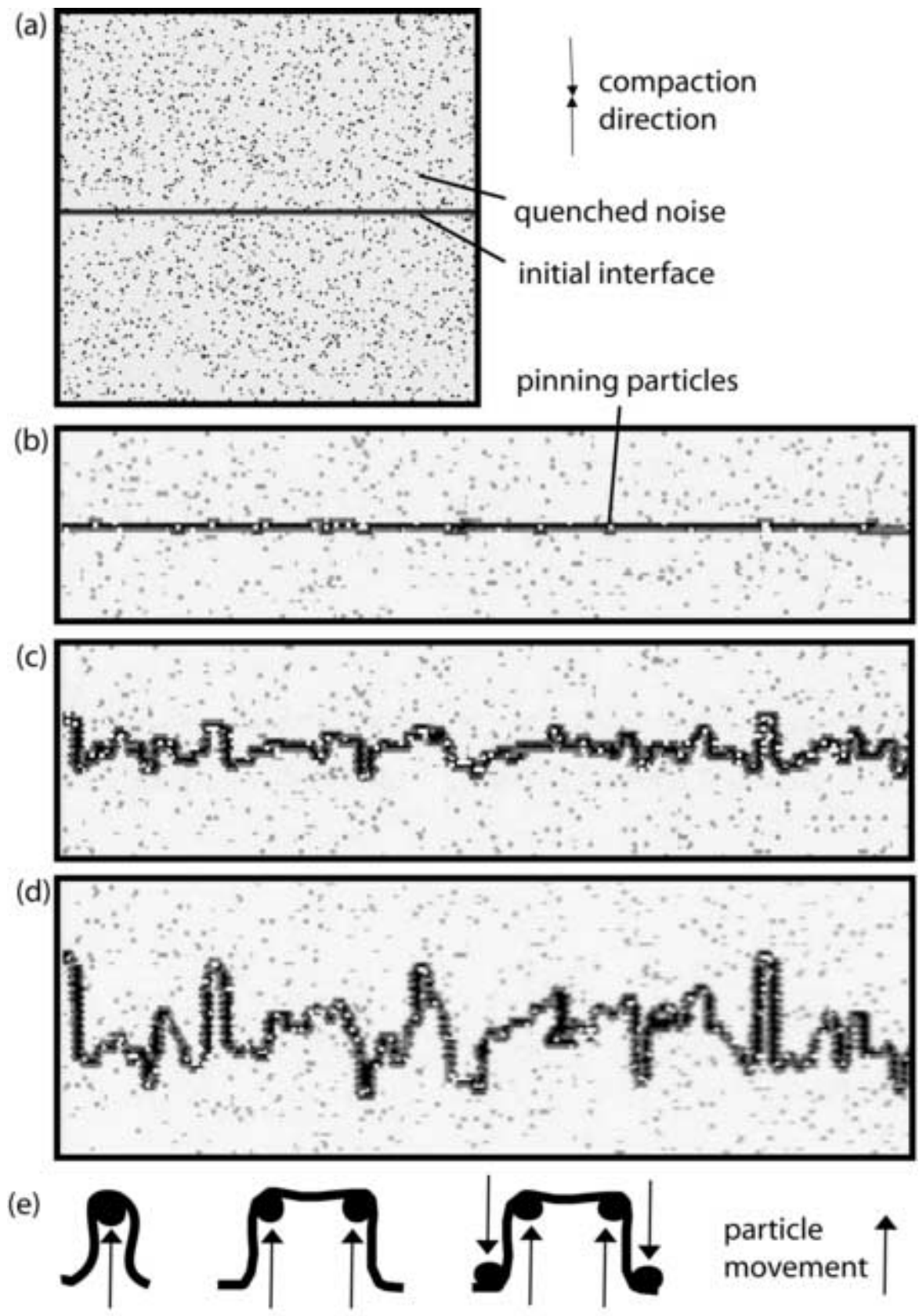

Fig. 4

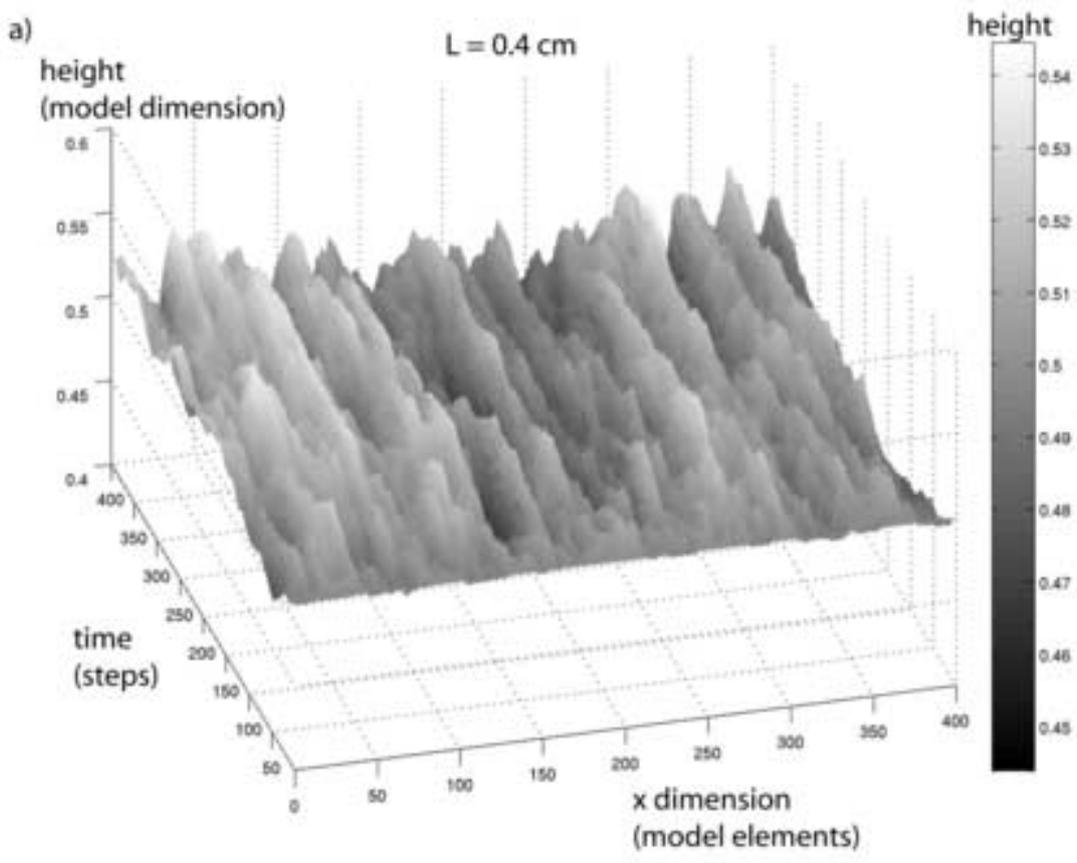

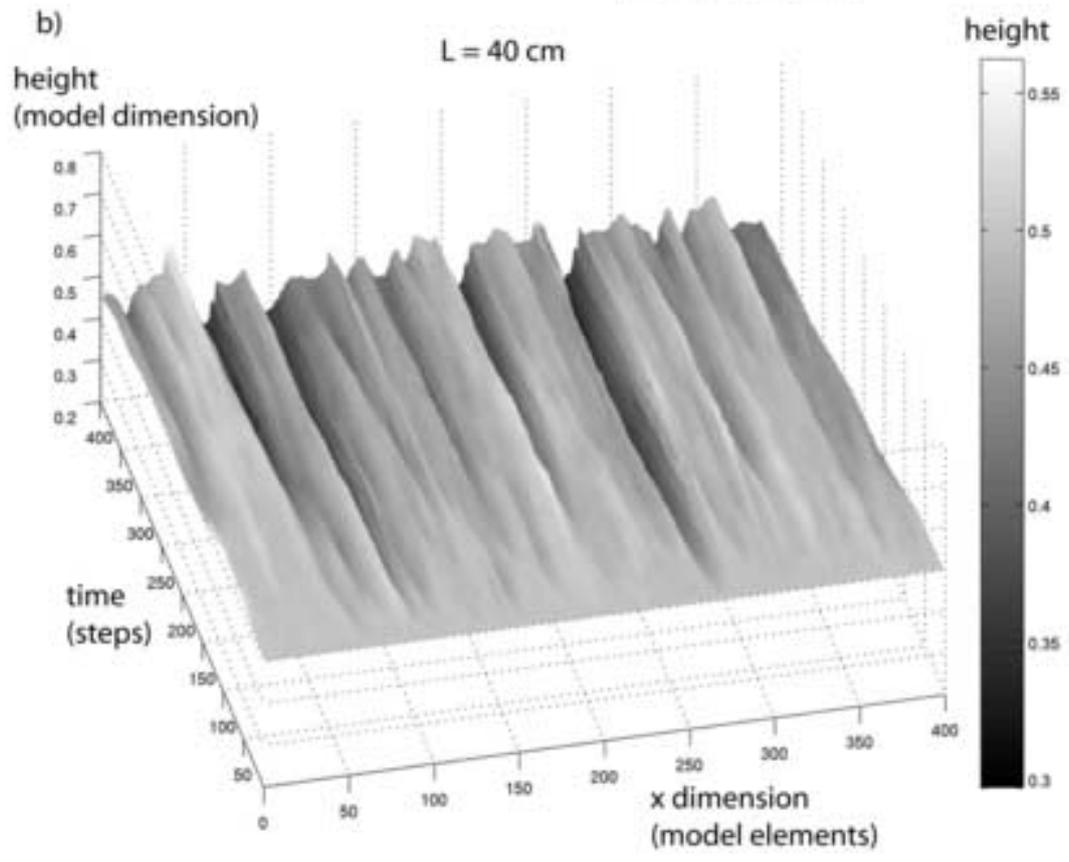

Fig. 5

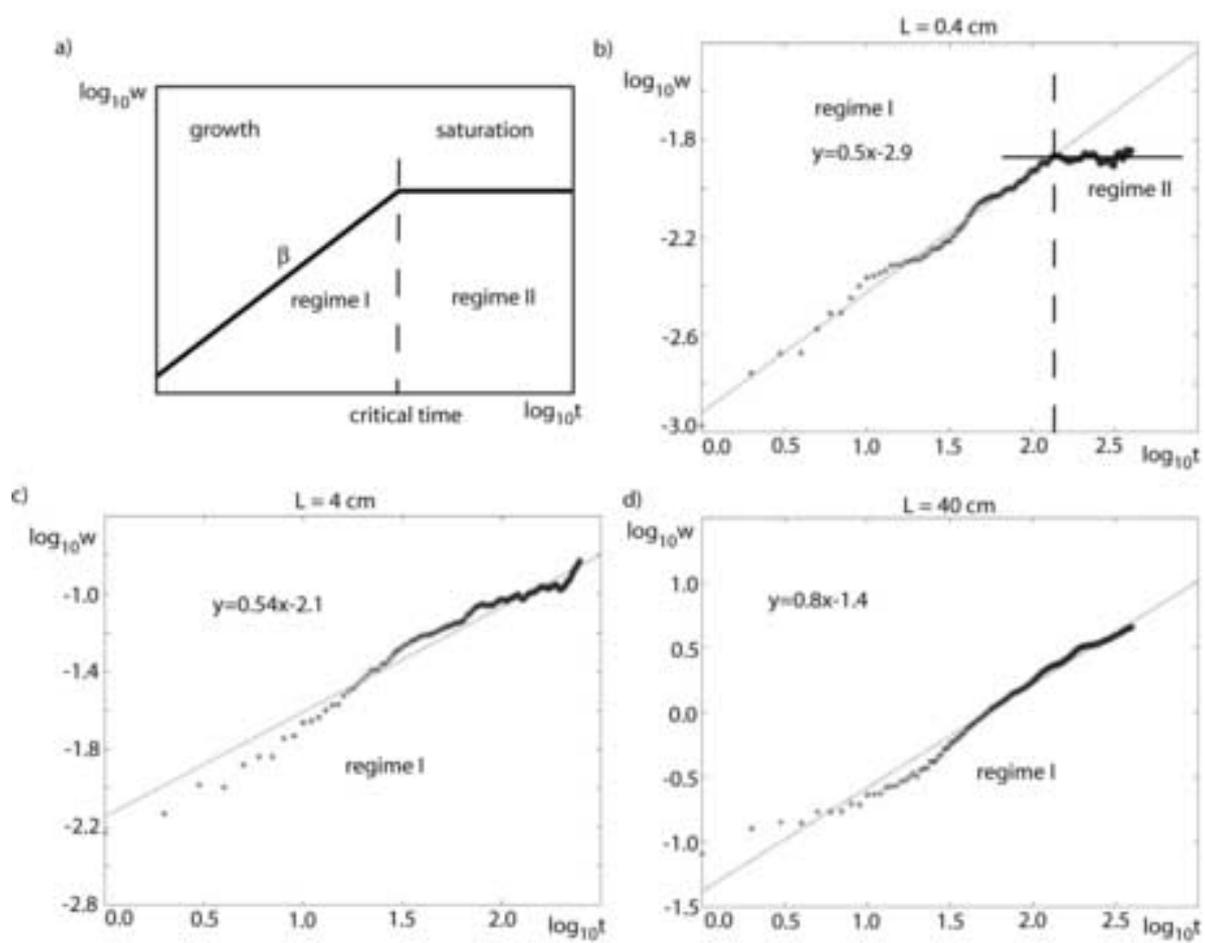

Fig. 6

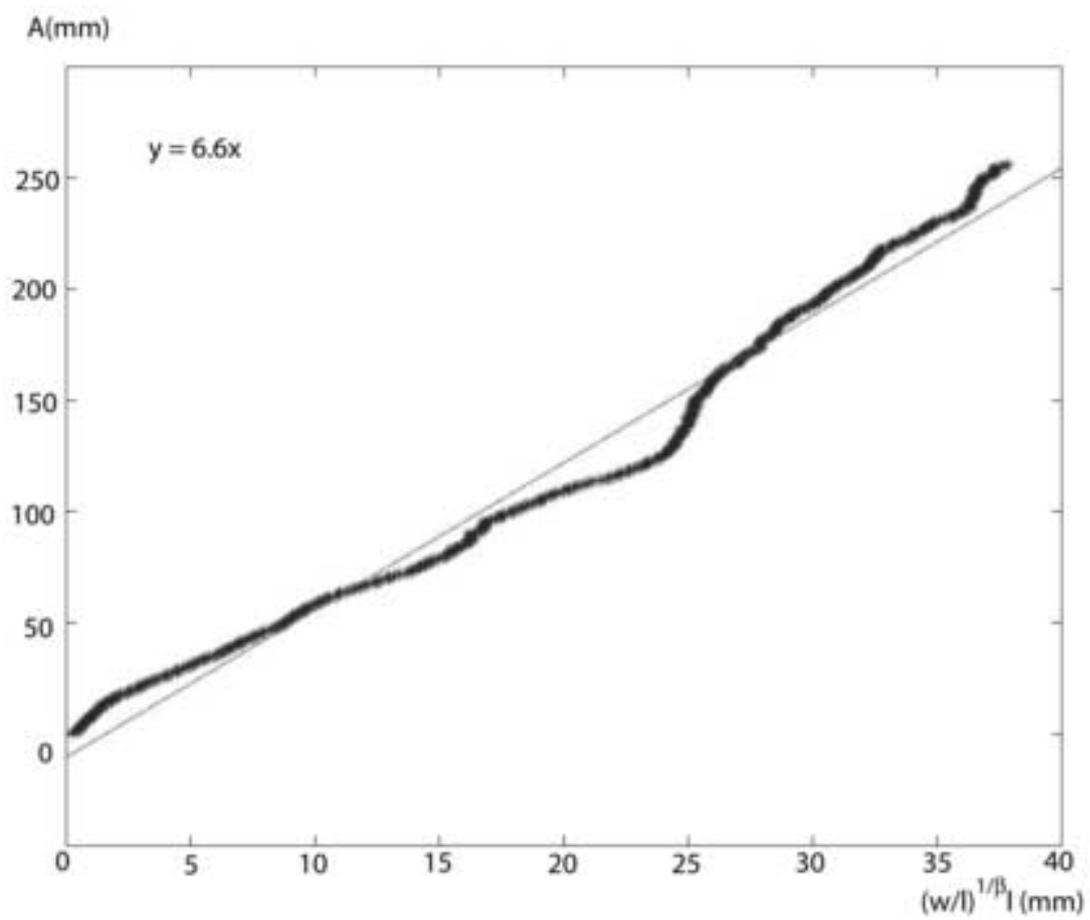

Fig. 7

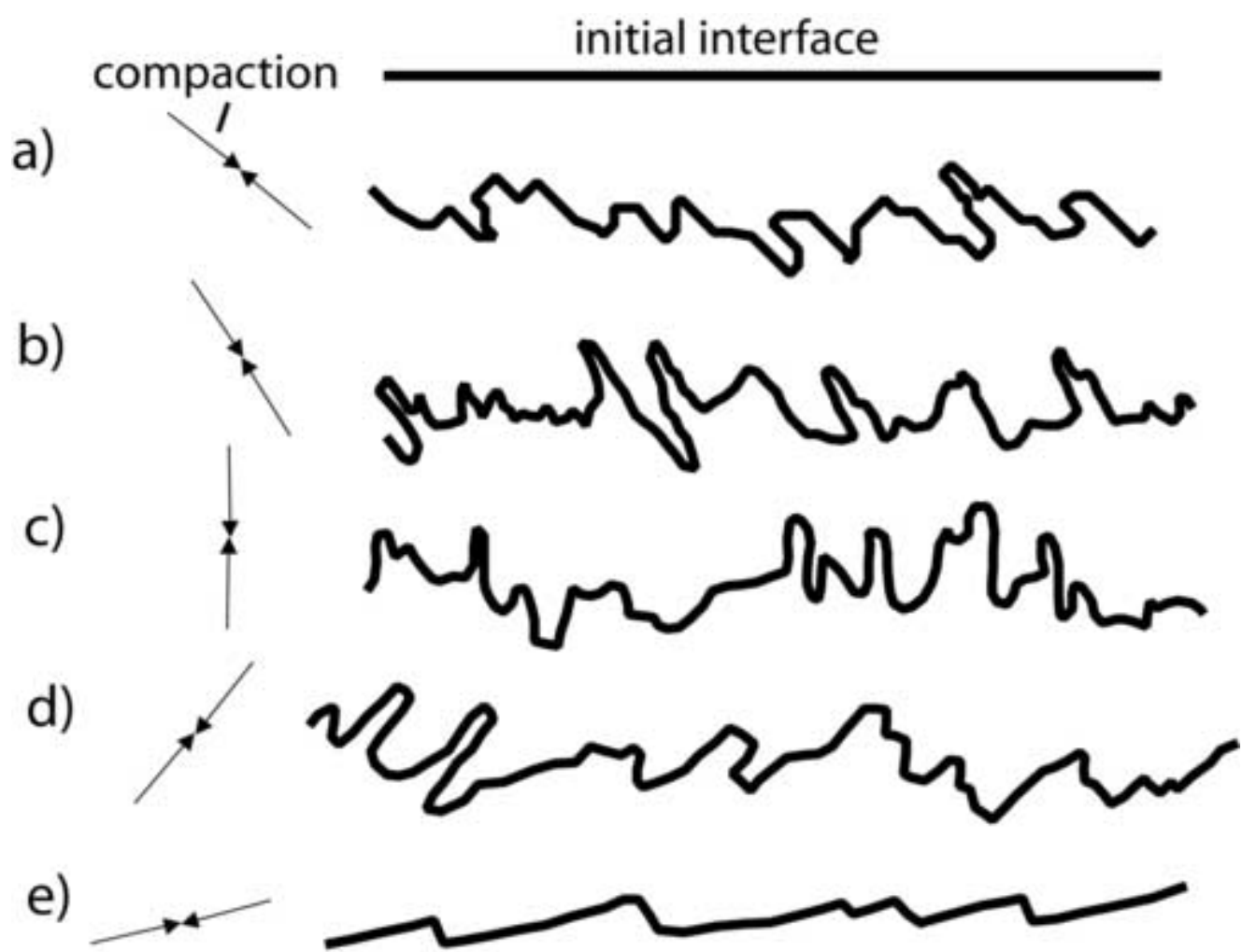

Fig. 8

**Figure**
**Click here to download high resolution image**

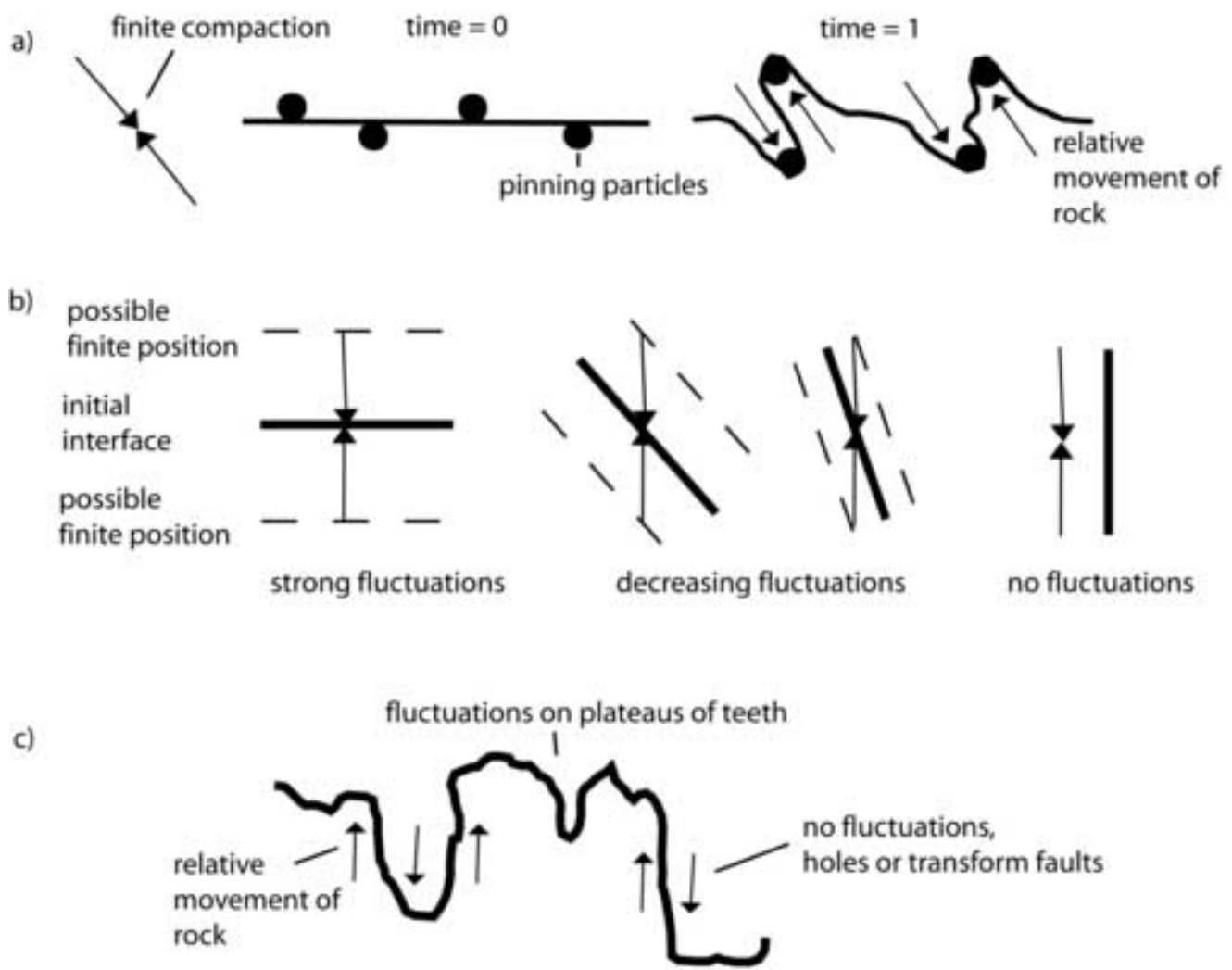

Fig. 9